\documentclass[aps,pra,twocolumn,groupedaddress,showpacs]{revtex4}

\usepackage{graphicx}
\newcommand{\rydn}{\textsf{n}}

\begin{document}


\title{Modeling many-particle mechanical effects of an interacting Rydberg gas}


\author{Thomas Amthor}
\email[]{thomas.amthor@physik.uni-freiburg.de}
\author{Markus Reetz-Lamour}
\author{Christian Giese}
\author{Matthias Weidem{\"u}ller}
\email[]{matthias.weidemueller@physik.uni-freiburg.de}
\affiliation{Physikalisches Institut, Universit{\"a}t Freiburg,
  Hermann-Herder-Str. 3, 79104 Freiburg, Germany}
\homepage[]{http://quantendynamik.physik.uni-freiburg.de}


\date{\today}

\begin{abstract}
 In a recent work [Phys. Rev. Lett. {\bf 98}, 023004 (2007)] we have
 investigated the influence of attractive van der Waals interaction on the
 pair distribution and Penning ionization dynamics of ultracold Rydberg
 gases. Here we extend this description to atoms initially prepared 
 in Rydberg states exhibiting repulsive interaction.
 We present calculations based on a Monte Carlo algorithm
 to simulate the dynamics of many atoms under the
 influence of both repulsive and attractive long-range interatomic forces.
 Redistribution to nearby states induced by black body radiation is taken into
 account, changing the effective interaction potentials. The model agrees with
 experimental observations, where the ionization rate is found to
 increase when the excitation laser is blue-detuned from the atomic resonance.
\end{abstract}


\maketitle



\begin{figure}
\includegraphics[scale=0.9,angle=0]{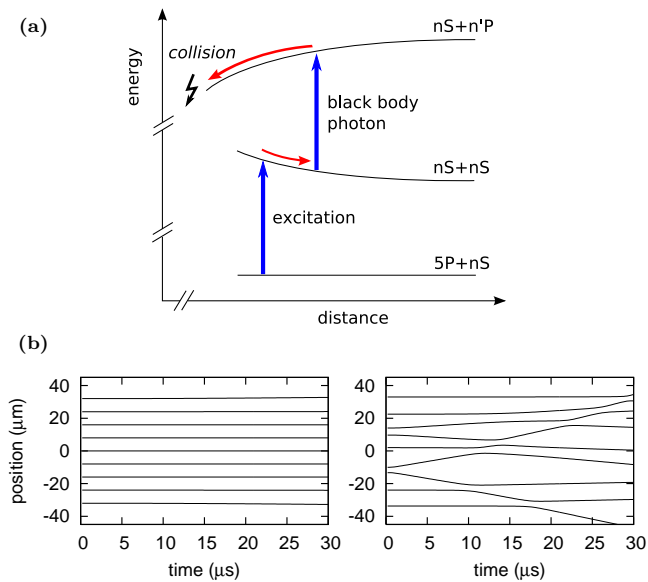}%
\caption{\label{fig:1dmodel}(Color online)
  (a) Rydberg atoms are excited in a repulsive
  potential. Black-body photons can transfer one atom
  to a dipole coupled state, allowing for attractive interaction.
  (b) A one-dimensional calculation of particle
  trajectories to illustrate the idea of many-particle repulsive
  interaction. When all particles have the same (large) initial distance from
  their neighbors, they will only be slightly repelled (left graph). When
  some close pairs exist in the beginning, there will repeatedly be atoms
  coming close to  each other over a long time scale (right graph).
  Here an average initial distance of 8\,$\mu$m and a van der Waals potential
  with $C_6=-10^{21}$\,au is assumed.}
\end{figure}

Ultracold Rydberg gases represent a unique system for studying atomic many-body
dynamics on experimentally observable time scales. Among many other
applications \cite{epjd06,jpb05}, these dilute and yet strongly
interacting gases are the starting point for the formation of ultracold plasmas
\cite{robinson2000,gallagher2003,pohl2003,walz-flannigan2004,robicheaux2005}.
While the cold ground state atoms can be considered to be fixed in space,
highly excited atoms are accelerated due to their long-range interaction potentials.
In a previous work we have shown how attractive long-range van der Waals
interaction influences the distribution of pair distances during excitation
and is responsible for the initial collisional ionization in an ultracold
Rydberg gas \cite{amthor2007}. Ions produced by the Penning ionization process
\begin{equation}
 A^{**} + A^{**} \rightarrow A^{*} + A^{+} + e^{-}
\end{equation}
were seen to appear earlier when the
excitation laser was red-detuned from the atomic resonance in a system with
purely attractive interactions. 
As explained in \cite{amthor2007}, 
a red-detuned laser preferentially excites close pairs,
which in turn collide on shorter time scales.
In that work the question remained, how collisional ionization can
be described in a system with only repulsive van der Waals
interaction (as is the case for Rydberg S states).
Experimental observation showed that the gas was ionizing on
longer time scales compared to the attractive case, and ions appeared earlier
when the excitation was blue-detuned \cite{amthor2007}.
Collisional ionization requires
attractive potentials, thus we must assume that some atoms are transferred to
other states after some time, thereby experiencing an attractive dipole-dipole
interaction with their neighbors. Fig. \ref{fig:1dmodel}a illustrates
this procedure.
Apart from state-changing collisions with electrons which already require some
free charges in the cloud \cite{walz-flannigan2004},
the main source for redistribution of
Rydberg states is black-body radiation \cite{gallagher1994}.
An experiment enforcing redistribution with microwave fields
has shown that the resulting attractive forces lead to
interaction-induced collisions \cite{li2005}.

However, a simple two-atom picture as shown in Fig. \ref{fig:1dmodel}a
with both atoms initially experiencing
repulsive forces is not sufficient to explain our observations,
since by the time a redistribution to an attractive potential takes place,
initially close atoms have already been accelerated too far away from
each other to return within observable time scales.
In order to describe the system more realistically, one has to take into
account the dynamics of many particles under the influence of repulsive
interactions.

Here we show how the experimental observations can be reproduced
by considering black-body redistribution and motion of many particles.
Fig. \ref{fig:1dmodel}b illustrates the idea of the many-body
aspect in a one-dimensional example:
When all particles at a given average density have the largest possible
(equal) initial distance from their neighbors,
they are only slightly repelled from each other (left graph).
Atoms redistributed to other states at any time need a
rather long time to be accelerated towards a neighbor, as the separations are
always large.
By contrast, when some close pairs exist in the beginning, as is the case
for blue-detuned excitation, there will constantly be atoms coming close
to each other -- even on a long time scale (right graph of
Fig. \ref{fig:1dmodel}b). Atoms redistributed to other states at a certain time
would then have the chance to collide with their neighbors earlier compared to
red-detuned excitation. The same idea holds for the three-dimensional case
implemented in our model described below.\\

\begin{figure*}
\includegraphics[scale=1]{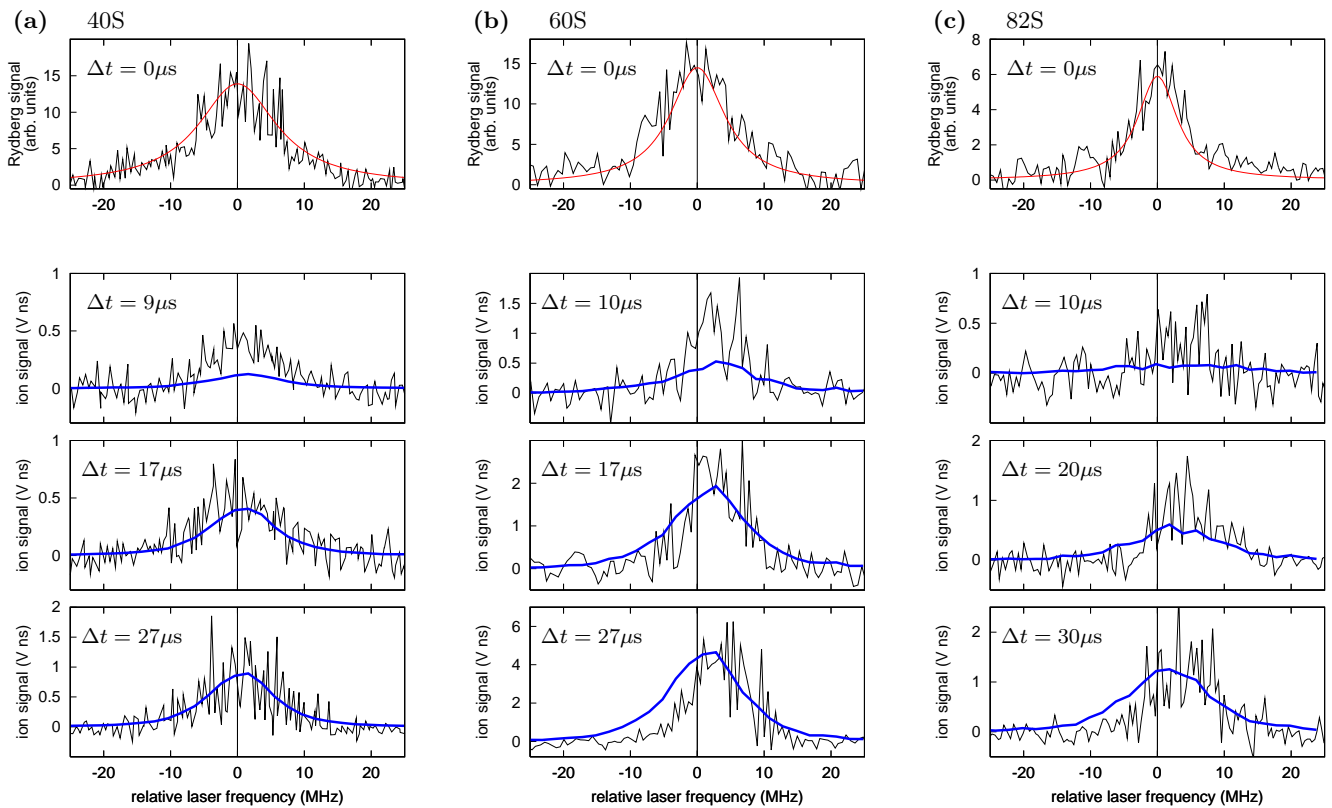}%
\caption{\label{fig:expvsmodel} (Color online)
  Expected collisional ionization when exciting
  in a repulsive potential. The upper graphs show the Rydberg signal directly
  after excitation, including a Lorentz fit to determine excitation fraction
  and saturation to be used in the model. 
  The other graphs show the ion signal detected after
  various delays. The corresponding simulated signals are plotted as bold
  (blue) lines.  (a) \rydn=40, (b) \rydn=60, (c) \rydn=82 
}
\end{figure*}

The experimental setup is identical to the one described in \cite{amthor2007}.
We trap $^{87}$Rb atoms in a standard
magneto-optical trap (MOT) at temperatures below 100\,$\mu$K and a peak
density on the order of $10^{10}$cm$^{-3}$. Excitation to Rydberg
states is accomplished by a two-photon excitation scheme, the two transitions
5S$_{1/2}\rightarrow$ 5P$_{3/2}$ and 5P$_{3/2}\rightarrow \rydn\ell$ being
realized with two cw laser systems at 780\,nm and 480\,nm, respectively.
An ultra-stable reference cavity is used to actively stabilize the
480-nm excitation laser.
The 780-nm laser illuminates the whole trapped atom cloud homogeneously,
while the 480-nm laser is focused with a waist of $\sim 37\,\mu$m at
the center of the MOT. 
Every 70\,ms the excitation laser is switched on for 200\,ns at a given
detuning and the gas can then evolve freely for a variable time $\Delta
t$. After that, an electric field ramp is applied 
to field ionize the Rydberg atoms. The ions are detected on a microchannel
plate detector. Ions already present before the field ramp is switched on
({\em i.e.} those produced by collisions) will reach the the detector earlier, while
Rydberg states ionize at a finite electric field, resulting in a delayed
detector signal. The two signals are recorded simultaneously by two boxcar
integrators. The Rydberg and ion signals have been measured for different
initial Rydberg states (40S, 60S, and 82S), all exhibiting purely
repulsive van der Waals interaction \cite{singer2005b}. 
 For each of the experiments we have verified that no
ions are produced during the excitation. While at \rydn=40 no significant
spectral asymmetry of the ion production rate has been observed, at \rydn=60
and \rydn=82 we see ions appear earlier on the blue-detuned side of the
resonance.
The measurements are displayed in Figure \ref{fig:expvsmodel}. The
graphs include Lorentzian fits to the Rydberg excitation line from which we
estimate the excitation fraction and saturation which is used in the
corresponding simulation. \\
\begin{table}
\caption{\label{tab:modelparameters}Model parameters. Rates are given in s$^{-1}$, interaction coefficients in atomic units.}
\begin{ruledtabular}
\begin{tabular}{cccccc}
\rydn & $R_{bb}$ &  $R_{em}$ &  $C_6$   & $C_3^{eff}$  & $C_3^{max}$      \\
\hline
40    &  12730              &  17325  & $-7.0\times 10^{18}$ & $2\times 10^{5}$ & $8.1\times 10^{5}$  \\
60    &   5680              &   4845  & $-1.0\times 10^{21}$ & $4\times 10^{6}$ & $4.5\times 10^{6}$  \\
82    &   3030              &   1853 & $-3.9\times 10^{22}$ & $3\times 10^{7}$ & $1.7\times 10^{7}$   \\
\end{tabular}
\end{ruledtabular}
\end{table}%

Our model considers the interaction-induced forces among a large number of
particles and calculates the motion of the atoms by numerically integrating
the equations of motion.
In the first part of the model, the Rydberg excitation process is
simulated.  5000 atoms are randomly placed in a cube to
represent a density of $10^{10}\,$cm$^{-3}$ corresponding to the center of 
the MOT. The iterative excitation procedure is similar to the
one used in Ref. \cite{amthor2007}. In each iteration step,
each atom $i$ has a certain probability for excitation or stimulated emission
\begin{equation}
 {\mathcal P}_i = a \, {\mathcal L}(\delta-V_{\mathrm{vdW}}(R_i)) \, e^{-2(x_i^2+y_i^2)/w^2} \, .
\end{equation}
$\mathcal L$ represents the laser line shape, determined by
the width of the intermediate state 5P (6\,MHz FWHM),
the widths of the two lasers (1--2\,MHz),
and the Fourier transform of the excitation pulse ($\sim$3\,MHz FWHM).
$\delta$ is the detuning from
resonance and $V_{\mathrm{vdW}}(R_i)$ is the interaction energy with the nearest
Rydberg neighbor. The last factor accounts for the Gaussian intensity
distribution of the excitation laser. The factor
$a=0.001$ is a small value to ensure a small number of excitations per
iteration step. The procedure is iterated several hundred times to yield the
actual percentage of excited atoms observed in the experiment (10--30\% at
$\delta=0$). For each atom, the distance $R_i$ to its nearest Rydberg neighbor
is recalculated after each excitation or stimulated emission event.
In the second part of the model, the interaction-induced motion of the
Rydberg atoms is calculated.
During excitation, the interaction between a pair of atoms is assumed to be of
van der Waals type,
\begin{equation}
 \label{eqn:vanderwaals}
 V_{\mathrm{vdW}}(R)=-\frac{C_6}{R^6}\; ,
\end{equation}
as all atoms are in a \rydn S state. 
The values for $C_6$ are calculated using a perturbative approach
\cite{singer2005b}.
Directly after excitation, the atoms are assumed to be at rest, and are then
free to move according to the forces caused by the interatomic
interaction. These dynamics are simulated by numerical integration of the
equations of motion, considering the interactions among all atoms.
Whenever the distance between two atoms becomes less than $4\rydn^2a_0$, these
atoms are assumed to collide and undergo Penning ionization
\cite{robicheaux2005,olson1979}.
While moving, the atoms can change their internal state in two ways:
\begin{enumerate}
 \item Black-body-induced redistribution to other Rydberg states. This process
   will mainly lead to nearby states, because of the higher transition matrix
   elements. The corresponding
   rates $R_{bb}$ are calculated following \cite{gallagher1994} and
   listed in Table \ref{tab:modelparameters}.
 \item Spontaneous emission. These atoms are transferred to low-lying states
   (due to the $\omega^3$ dependence of the Einstein $A$ coefficient)
   and do not show any significant long-range interactions with other
   atoms any more. They are thus removed from the simulation. The spontaneous
   emission rates $R_{em}$, calculated according to \cite{gallagher1994}, are
   comparable to the black-body redistribution and must be considered to
   reproduce the time development of the experimental ion signal (see Table
   \ref{tab:modelparameters}).
\end{enumerate}
Direct photo-ionization by black-body radiation is also included in the model,
but the rates for this process are small ($<200$\,Hz) \cite{beterov2007} and
are exceeded by the collision rates after the first few $\mu$s.
Two different types
of interaction potentials are considered in the equations of motion: The
(repulsive) interaction between atoms in the initial state is described by
the potential in Eq. (\ref{eqn:vanderwaals}).
Attractive potentials are introduced via the dipole-dipole interaction of an
atom in the initial state (\rydn S) with an atom in a redistributed state
(\rydn$'$P), given by
\begin{equation}
 \label{eqn:dipoledipole}
 V_{\mathrm{dd}}(\vec R)=\frac{\vec \mu_1\vec \mu_2}{R^3}-3\frac{(\vec \mu_1\vec
   R) (\vec\mu_2\vec R)}{R^5}\; ,
\end{equation}
with $|\vec \mu_1|=|\vec \mu_2|=\mu_{eff}$ and an effective coefficient
$C_3^{eff}=\mu_{eff}^2$, representing the average over different electronic
states after redistribution, and random angular dipole alignment.
The largest possible transition dipole moment
$\mu_{max}$ is found between \rydn S and \rydn P (\rydn$'$=\rydn), 
which leads to a maximum effective interaction strength
$C_3^{max} = \mu_{max}^2$.
We have simulated the ionization dynamics for the three different
principal quantum numbers investigated experimentally.
The parameters used in the model are listed
in Table \ref{tab:modelparameters}. The effective attractive
interaction strengths $C_3^{eff}$ are chosen to reproduce the time
development of the data best.
The bold lines in Fig. \ref{fig:expvsmodel} show the results of the
simulation, averaged over 100 runs.
In order to compare the line shapes, the
simulated traces have been scaled to the experimental data. The
scaling factor is the same for all traces belonging to the same principal
quantum number \rydn.
To achieve satisfactory overlap of the curves, we used different scaling
factors for the three values of \rydn. We attribute this to a
number of effects which are not included in the model:
First, the simulation is performed for a constant density, while in
the experiment the excitation volume also extends over regions with
lower density. The excitation fraction in the center of the cloud
can only be roughly estimated, and the excitation is saturated to
different amounts for the different measurements.
Even more importantly, any secondary effects of the collision products, such as
local electric fields and state-changing collisions \cite{walz-flannigan2004}, 
as well as avalanche ionization effects, are neglected in the model.
For higher principal quantum numbers, a considerable amount of atoms may
therefore experience a mixing of states in addition to the black-body
redistribution. For this reason we expect the the ion production to be
underestimated for higher \rydn, and accordingly the $C_3^{eff}$ to be
overestimated.
In fact, for the 40S state the simulated ionization fraction at 30\,$\mu$s
(3.0\%) corresponds to the experimental estimate of 4--8\%. At higher \rydn\
the ionization fraction is systematically underestimated, for 82S we obtain
1.5\% ions from the model, compared to around 15\% in the experiment. (The
experimental fractions are subject to an uncertainty of about a factor of two.)
At the same time the attractive interaction is systematically overestimated as
\rydn\ increases (see Table \ref{tab:modelparameters}). If only black-body
redistribution rates were involved in the dynamics, $C_3^{eff}$ should always
be less than $C_3^{max}$, the maximum interaction strength with an
energetically close P state. This suggests that
other redistribution processes and the influence of collision products become
increasingly important.

Despite these limitations, the qualitative behavior of the dynamics of the
system is very well reproduced: The maximum ionization rate appears on the
blue-detuned side of the atomic resonance, and this spectral shift increases
towards higher \rydn, as the interaction-induced variation in the pair distance
distribution becomes more distinct. 
The spectral shift of the maximum ion signal is mainly determined by the
excitation process, and is insensitive to the choice of attractive
potentials. It is best reproduced for delay times below 25\,$\mu$s.
\\

We have presented a Monte Carlo excitation model in combination with
a numerical calculation of particle trajectories to explain the ionization
behavior of a Rydberg gas initially prepared in a state with repulsive van
der Waals interaction. In order to reproduce 
measurements showing an increased ionization rate on the blue-detuned wing of
the excitation line, it is necessary to consider both black-body-induced
redistribution of states and the motion of many atoms, while two-atom models
fail to reproduce the experimental data. For increasing principal quantum
number \rydn, other redistribution processes and the influence of free charges
become apparent.
These results contribute to a more complete
picture of the processes triggering ionization and plasma formation in
ultracold Rydberg gases.\\



We thank T. Gallagher for valuable discussions and for bringing the importance
of black body radiation to our attention.
The project is supported in part by the Landesstiftung
Baden-W{\"u}rttemberg in the framework of the ``Quantum
Information Processing'' program, and a grant from the
Ministry of Science, Research and Arts of Baden-
W{\"u}rttemberg (Az: No. 24-7532.23-11-11/1).




\end{document}